\documentclass[12pt, a4paper]{article}
\usepackage{amsmath}

\begin{document}

\title{Hydrodynamic reductions of multi-dimensional dispersionless PDEs: the test for integrability}
\author{E.V. Ferapontov  \footnote{Center for Nonlinear Studies, Landau Institute for Theoretical Physics, Kosygina 2, Moscow, 117940, Russia.} \  and K.R. Khusnutdinova
   \footnote{On the leave from: Institute of Mechanics, Ufa Branch of
the Russian Academy of
   Sciences, Karl Marx Str. 6, Ufa, 450000, Russia.}}
   \date{}
   \maketitle
   \vspace{-7mm}
\begin{center}
Department of Mathematical Sciences \\ Loughborough University \\
Loughborough, Leicestershire LE11 3TU \\ United Kingdom \\[2ex]
e-mails: \\[1ex] \texttt{E.V.Ferapontov@lboro.ac.uk}\\
\texttt{K.Khusnutdinova@lboro.ac.uk}
\end{center}

\bigskip

\begin{abstract}
A $(d+1)$-dimensional dispersionless PDE is said to be integrable  if its $n$-component hydrodynamic reductions
are locally parametrized by $(d-1)n$ arbitrary functions of  one variable. Given a PDE which does not pass 
the integrability test, the method of hydrodynamic reductions allows one to effectively reconstruct  additional differential constraints
which, when added to the equation, make it an integrable system in fewer dimensions (if consistent).



\end{abstract}

\newpage

\section{Introduction}

We address the problem of integrability of
multi-dimensional dispersionless PDEs of the form
\begin{equation}
F(u, u_i, u_{ij})=0
\label{1}
\end{equation}
where $u$ is a (vector-) function  of $d+1$ independent variables. For definiteness, let us consider $(3+1)$-dimesional PDEs in four independent variables  $t, x, y, z$. 
Equations  of this type naturally arise in mechanics, mathematical physics, general relativity and differential geometry. Let us look for exact solutions of  (\ref{1}) of the form
${ u}={ u}(R^1, ..., R^n)$  where the {\it Riemann invariants}
$R^1, ..., R^n$ solve a triple of
commuting diagonal systems
\begin{equation}
R^i_t=\lambda^i(R)\ R^i_x, ~~~~ R^i_y=\mu^i(R)\ R^i_x, ~~~~ R^i_z=\eta^i(R)\ R^i_x.
\label{R}
\end{equation}
Notice that the number of Riemann invariants is allowed to be
arbitrary!  Thus, the original  multi-dimensional equation (\ref{1}) is
decoupled into a collection of commuting  $(1+1)$-dimensional systems in Riemann invariants. 
In some cases one first needs to rewrite the original PDE (\ref{1}) in a quasilinear form to make the method work: see  examples in Sect.2.
Solutions of this type, known as nonlinear interactions of $n$ planar
simple waves, were  investigated in gas dynamics and
magnetohydrodynamics  \cite{ Burnat3,  Perad2}. Later, they 
appeared in the context of the dispersionless KP hierarchy
\cite{Gibb94, GibTsa96, GibTsa99, GuMaAl, Ma, MaAl, Boyarsky}. We will call a
multi-dimensional equation {\it integrable} if it possesses
`sufficiently many' $n$-component reductions of the form (\ref{R})
for arbitrary $n$ (the precise definition follows). Exact solutions arising within this approach
 can be viewed as   dispersionless
analogues of  $n$-gap solutions. 

Partial classification results of $(2+1)$-dimensional  integrable systems of hydrodynamic type (the integrability is understood in the above sense) were obtained in our 
recent publications \cite{Fer3, Fer4}.  Integrable second order PDEs of the form $u_{tt}=f(u_{xx}, u_{xt}, u_{xy})$ were classified in \cite{Pavlov1}.
Particularly interesting examples arise in the theory of self-dual and Einstein-Weyl spaces \cite{Plebanski, dun1, dun3, Fer, Fer2}, in the context of the Dirichlet boundary problem in multi-connected domains  \cite{Kr3}, and the Whitham averaging procedure (in particular, the dispersionless limit) applied to $(2+1)$-dimensional solitonic PDEs  \cite{Kr1, Kr2, Zakharov}.

We recall, see \cite{Tsarev}, that the requirement of  
commutativity of the flows (\ref{R})
is equivalent to the following restrictions on their characteristic speeds:
\begin{equation}
\frac{\partial_j\lambda
^i}{\lambda^j-\lambda^i}=\frac{\partial_j\mu^i}{\mu^j-\mu^i}=\frac{\partial_j\eta^i}{\eta^j-\eta^i},
\label{comm}
\end{equation}
$i\ne j, \  \partial_j=\partial/\partial_{ R^j}$; no summation! Once these conditions are met, the general solution
of (\ref{R}) is given by the
implicit  `generalized hodograph'  formula \cite{Tsarev}
\begin{equation}
v^i(R)=x+\lambda^i(R)\ t+\mu^i(R) \ y+\eta^i(R)\ z, ~~~ i=1, ..., n,
\label{hod}
\end{equation}
where $v^i(R)$ are  characteristic speeds of the general flow
commuting with (\ref{R}), that is, the general solution of the linear
system
\begin{equation}
\frac{\partial_jv^i}{v^j-v^i}=\frac{\partial_j\lambda
^i}{\lambda^j-\lambda^i}=\frac{\partial_j\mu^i}{\mu^j-\mu^i}=\frac{\partial_j\eta^i}{\eta^j-\eta^i}.
\label{comm1}
\end{equation}
Substituting ${ u}(R^1, ..., R^n)$ into (\ref{1}) and using
(\ref{R}) one  arrives at an over-determined system for $\lambda^i(R), \mu^i(R), \eta^i(R)$ and $u(R)$ as functions of the Riemann invariants $R^i$. 
This system implies, in particular, that the characteristic speeds $\lambda^i, \mu^i$ and $\eta^i$ satisfy an algebraic relation which can be interpreted 
as the dispersion relation for the system (\ref{1}).

One can show that the requirement of   existence of nontrivial
3-component reductions is already
sufficiently  restrictive  implying, in particular,  the
existence of  $n$-component reductions for arbitrary $n$ \cite{Fer3}. This
phenomenon is similar to the well-known three-soliton condition in
the Hirota bilinear approach \cite{Hietarinta2}
(recall that two-soliton solutions exist for arbitrary PDEs
transformable to Hirota's bilinear form and, therefore, cannot detect
the integrability),
and the condition of three-dimensional consistency in
the classification of discrete integrable systems on quad-graphs \cite{Adler}.
One can show that the maximum number of $n$-component reductions a (d+1)-dimensional PDE 
may possess is parametrized, modulo changes of variables $R^i\to
f^i(R^i)$,  by $(d-1)n$ arbitrary functions of one variable. Therefore, we propose the
following

{\bf Definition.} {\it A (d+1)-dimensional PDE is said to be integrable if its n-component  reductions are locally parametrized by $(d-1)n$ arbitrary functions of one variable.}

One of the most important examples of PDEs  in four dimensions which are integrable in this sense is the `second heavenly' equation, 
$$
\theta_{tx}+\theta_{zy}+\theta_{xx}\theta_{yy}-\theta^2_{xy}=0,
$$
descriptive of self-dual Einstein spaces \cite{Plebanski} (see Example 2 of Sect.2). Its equivalent 
`first heavenly' form was discussed  in the recent publication  \cite{Fer2} where it was shown that $n$-component reductions  are parametrized by $2n$ arbitrary functions of a single variable. It would be interesting to compare these reductions with the solitonic reductions of \cite{dun2}. The requirement of existence of $n$-component reductions parametrized by $2n$ arbitrary functions of one variable appears to be very strong, indeed, the heavenly equation is the only nonlinear PDE of the form
$$
\theta_{tx}+\theta_{zy}=f(\theta_{xx}, \theta_{xy}, \theta_{yy}),
$$
which passes the integrability test (see the Appendix).

An interesting six-dimensional integrable generalization of the heavenly equation,
$$
\theta_{t\tilde t}+\theta_{z\tilde z}+\theta_{tx}\theta_{zy}-\theta_{ty}\theta_{zx}=0,
$$
arises in the context of $sdiff(\Sigma^2)$  self-dual Yang-Mills equations  \cite{Przanovski}.  Its $n$-component  reductions are parametrized by $4n$ arbitrary functions of a single variable (Example 5 of Sect.3). We would like to thank M. Dunajski for drawing our attention to this equation.

Another  integrable example is the six-dimensional system
$$
m_{ t}=n_{ x}+nm_{ r}-mn_{ r},  ~~~~ n_{ z}=m_{ y}+mn_{ s}-nm_s,
$$
see Sect.3. Its $n$-component reductions depend on $4n$ arbitrary functions of one variable. Under the additional constraints $m_r=n_r=0, \ s=x, \ z=t$, this system is descriptive of hyperCR Einstein-Weyl structures  \cite{dun0}; it was investigated in a series of recent publications  \cite{Pavlov, dun3, Alonso}.

In the Example 3 of Sect.2 we apply our method to the four-dimensional PDE
 $$
 F_{xz}=F_{xx}F_{xy}+F_{yt}
$$
which is the second flow of the dispersionless KP (dKP) hierarchy
 $$
\begin{array}{c}
F_{xt}=\frac{1}{2}F_{xx}^2+F_{yy},~~\\
\ \\
 F_{xz}=F_{xx}F_{xy}+F_{yt},\\
 .........................
\end{array}
$$
see, e. g., \cite{Carroll}. It is demonstrated that, considered  separately, this equation is {\it not} integrable (as a four-dimensional PDE), the fact which is not at all surprising. What is more important, the method of hydrodynamic reductions allows one to effectively reconstruct the differential constraint $F_{xt}=\frac{1}{2}F_{xx}^2+F_{yy}$ which, when added to the equation, generates the $(2+1)$-dimensional dKP hierarchy.

Although the method of hydrodynamic reductions provides an infinity of (implicit) solutions parametrised by arbitrarily many functions of one variable, 
the question of solving the initial value problem for integrable multi-dimensional dispersionless PDEs remains open.  A detailed investigation of the behavior and singularity structure of solutions describing nonlinear interactions of planar simple waves  is beyond the scope of this paper.

\section{Examples}

In this section we list some  examples of
multi-dimensional  PDEs  which are
integrable in the  sense of hydrodynamic reductions.

{\bf Example 1.} Let us consider the first two flows of the dispersionless KP hierarchy, 
\begin{equation}
\begin{array}{c}
F_{xt}=\frac{1}{2}F_{xx}^2+F_{yy}, \\
\ \\
 F_{xz}=F_{xx}F_{xy}+F_{yt},
\end{array}
\label{dkp}
\end{equation}
which, in the new variables 
$$F_{xx}=u, ~ F_{xy}=v, ~ F_{yy}=w, ~ F_{yt}=s, ~ F_{xt}=\frac{1}{2}u^2+w, ~ F_{xz}=uv+s,$$
assume the quasilinear form
\begin{equation}
\begin{array}{c}
u_y=v_x, ~~ v_y=w_x, ~~ v_t=s_x, ~~ w_t=s_y, ~~ u_t=(\frac{1}{2}u^2+w)_x, ~~ u_z=(uv+s)_x, \\
\ \\
 v_t=(\frac{1}{2}u^2+w)_y, ~~ v_z=(uv+s)_y, ~~ s_x=(\frac{1}{2}u^2+w)_y, ~~ (\frac{1}{2}u^2+w)_z=(uv+s)_t.
\end{array}
\label{dkp1}
\end{equation}
Notice that $u$ satisfies the  dispersionless KP equation, 
$$
(u_t-uu_x)_x=u_{yy}.
$$
Looking for reductions in the form $u=u(R^1, ..., R^n), $ $ v=v(R^1, ..., R^n), $ $ w=w(R^1, ..., R^n), $ $s=s(R^1, ..., R^n)$
where the Riemann invariants $R^i$ satisfy  (\ref{R}), and substituting into (\ref{dkp1}),  one readily obtains
\begin{equation}
\partial_i v=\mu^i \partial_i u, ~~~ \partial_i s=\lambda^i \mu^i \partial_i u, ~~~ \partial_i w=(\mu^i)^2 \partial_i u,
\label{dkp2}
\end{equation}
along with the dispersion relations
\begin{equation}
\lambda^i=u+(\mu^i)^2, ~~~ \eta^i=v+2u\mu^i+(\mu^i)^3.
\label{dispdkp}
\end{equation}
The compatibility condition
$\partial_i\partial_jv=\partial_j\partial_iv$ implies
\begin{equation}
\partial_i\partial_ju=\frac{\partial_j\mu^i}{\mu^j-\mu^i}\partial_iu+\frac{\partial_i\mu^j}{\mu^i-\mu^j}\partial_ju,
\label{dkp3}
\end{equation}
while the commutativity condition (\ref{comm}) results in
\begin{equation}
\partial_j\mu^i=\frac{\partial_j u}{\mu^j-\mu^i}.
\label{dkp4}
\end{equation}
The substitution of (\ref{dkp4}) into (\ref{dkp3}) implies the
Gibbons-Tsarev system for $u(R)$ and $\mu^i(R)$,
\begin{equation}
\partial_j\mu^i=\frac{\partial_j u}{\mu^j-\mu^i}, ~~~
\partial_i\partial_ju=2\frac{\partial_iu\partial_ju}{(\mu^j-\mu^i)^2},
\label{dkp5}
\end{equation}
$i\ne j$, which was first derived in \cite{GibTsa96, GibTsa99} in the
context of hydrodynamic reductions of
Benney's moment equations.
For any solution $\mu^i,  u$ of the system (\ref{dkp5}) one can
reconstruct $\lambda^i, \eta^i$ and $v, s, w$ by virtue of the relations (\ref{dispdkp}) and (\ref{dkp3}), which are automatically consistent.
 The general solution of the system (\ref{dkp5})
depends, modulo reparametrizations  $R^i\to f^i(R^i)$, on $n$
arbitrary functions of one variable, thus manifesting the fact that   PDEs (\ref{dkp}) constitute a (2+1)-dimensional integrable system.

{\bf Example 2.} The so-called second heavenly equation,
\begin{equation}
\theta_{tx}+\theta_{zy}+\theta_{xx}\theta_{yy}-\theta^2_{xy}=0,
\label{h1}
\end{equation}
is descriptive of self-dual Ricci-flat metrics \cite{Plebanski}. Introducing the variables $\theta_{xx}=u, $ $ \theta_{xy}=v, $ $ \theta_{yy}=w, $ $ \theta_{tx}=p, $ $ \theta_{zy}=v^2-uw-p$, one can rewrite 
(\ref{h1}) in a quasilinear form,
\begin{equation}
\begin{array}{c}
u_y=v_x, ~~~ u_t=p_x, ~~~ v_y=w_x, ~~~ v_t=p_y, \\
\ \\
v_z=(v^2-uw-p)_x, ~~~ w_z=(v^2-uw-p)_y.
\end{array}
\label{h2}
\end{equation}
Hydrodynamic reductions are sought in the form $u=u(R^1, ..., R^n),$ $v=v(R^1, ..., R^n),$ $w=w(R^1, ..., R^n),$ $p=p(R^1, ..., R^n)$  where the Riemann invariants $R^1, ..., R^n$  solve a triple of commuting hydrodynamic type systems (\ref{R}).
The substitution into (\ref{h2}) implies 
\begin{equation}
\partial_i p=\lambda^i\partial_i u, ~~~ \partial_i v=\mu^i\partial_i u, ~~~ \partial_i w=(\mu^i)^2\partial_i u,
\label{h4}
\end{equation}
along with the dispersion relation
\begin{equation}
\lambda^i=2v\mu^i-w-u(\mu^i)^2-\mu^i\eta^i.
\label{h5}
\end{equation}
Substituting  $\lambda^i$ into the commutativity conditions (\ref{comm}),
and taking into account that the compatibility conditions for the relations $\partial_i p=\lambda^i\partial_i u, \ \partial_i v=\mu^i\partial_i u$ imply
$$
\partial_i\partial_ju=\frac{\partial_j\mu^i}{\mu^j-\mu^i}\partial_iu+\frac{\partial_i\mu^j}{\mu^i-\mu^j}\partial_ju,
$$
one arrives at the following system:
\begin{equation}
\begin{array}{c}
\partial_j\mu^i=\frac{(\mu^j-\mu^i)^2}{\eta^j-\eta^i+u(\mu^j-\mu^i)} \ \partial_ju, ~~~~
\partial_j\eta^i=\frac{(\mu^j-\mu^i)(\eta^j-\eta^i)}{\eta^j-\eta^i+u(\mu^j-\mu^i)} \ \partial_ju, \\
\ \\
\partial_i\partial_j u=2\frac{\mu^j-\mu^i}{\eta^j-\eta^i+u(\mu^j-\mu^i)} \ \partial_iu\partial_ju.
\end{array}
\label{h6}
\end{equation}
Solving  equations (\ref{h6}) for $\mu^i,  \eta^i$ and $u$, determining $\lambda^i$ from (\ref{h5}) and calculating $p, v, w$ from  equations (\ref{h4}) 
(which are automatically compatible by virtue of (\ref{h6})), one obtains the general $n$-component hydrodynamic reduction of the heavenly equation.
Moreover, the commutativity conditions  will also be satisfied identically.

We emphasize that the system (\ref{h6}) is in involution and its general solution depends on $3n$ arbitrary functions of one variable.
Indeed, one can arbitrarily prescribe the restrictions of $\mu^i$ and $\eta^i$ to the $R^i$-coordinate line. This  gives $2n$ arbitrary functions. Moreover, one can arbitrarily 
prescribe the restriction of $u$ to each of the coordinate lines, which provides extra $n$  arbitrary functions. However, since  reparametrizations 
$R^i\to f^i(R^i)$ leave the system (\ref{h6}) invariant, one concludes that  general n-component reductions are locally parametrized by $2n$ arbitrary functions of one variable.
This supports the evidence that the heavenly equation (\ref{h1}) is a true four-dimensional integrable PDE.

Obviously, the same method  applies to other equivalent forms of the heavenly equation (\ref{h1}). For instance, hydrodynamic reductions of the  first heavenly equation
$$
\Omega_{xy}\Omega_{zt}-\Omega_{xt}\Omega_{zy}=1
$$
were investigated in detail in \cite{Fer2}. Another possibility is to work with the evolutionary form \cite{Grant} of the heavenly equation,
$$
\psi_{tt}=\psi_{xy}\psi_{zt}-\psi_{xt}\psi_{zy}.
$$
In both cases one can derive  analogues of  equations  (\ref{h6})  which, although involutive, look somewhat more complicated.

The requirement of  existence of    $n$-component  reductions (parametrized by $2n$ arbitrary functions of one variable) is  very strong indeed: as demonstrated in the Appendix, the heavenly equation (\ref{h1}) is the only nonlinear PDE of the form
$$
\theta_{tx}+\theta_{zy}+f(\theta_{xx}, \theta_{xy}, \theta_{yy})=0,
$$
which passes the integrability test.

{\bf Remark.} The heavenly equation (and  equivalent forms thereof) belong to the class of {\it special} Monge-Amp\'ere equations which can be defined as follows.
Consider a function $u(x^1, ... x^k)$ and introduce a  $k\times k$ symmetric matrix  $U=\vert u_{ij}\vert$ of its second partial derivatives. A special Monge-Ampere equation with constant coefficients is a PDE of the form
$$
M_0+M_1+...+M_k=0
$$
where $M_l$ is a constant-coefficient linear combination of all distinct $l\times l$ minors of  $U$, $0\leq l\leq k$. Here, for instance, $M_0$ is a constant, $M_k= det~U= Hess~u$, etc. Equivalently, this PDE can be obtained by equating to zero a constant-coefficient  $k$-form in $2k$ variables $x^i, u_i$. It is an interesting problem to classify integrable PDEs within this class, in particular, for $k=4$. We emphasize that the case $k=3$ is understood completely: one can show that, for $k=3$, any special Monge-Amp\'ere equation is either linearizable by a contact transformation
(in this case it is automatically integrable by the method of hydrodynamic reductions), or contact equivalent to either of the three nondegenerate forms \cite{Lychagin, Banos},
$$
Hess~u=1, ~~~  Hess~u=u_{11}+u_{22}+u_{33}, ~~~ Hess~u=u_{11}+u_{22}-u_{33}.
$$
We have verified directly  that these three PDEs are {\it not} integrable  by the method of hydrodynamic reductions.

{\bf Example 3.} Let us consider the second flow of the dispersionless KP hierarchy, 
\begin{equation}
F_{xz}=F_{xx}F_{xy}+F_{yt},
\label{dkp02}
\end{equation}
see Example 1. The question is:  should it be regarded as a four-dimensional integrable PDE?  We will see that the answer to this question is negative, moreover, the method of hydrodynamic reductions applied to this equations  reconstructs  additional differential constraints which, when added to (\ref{dkp02}), generate the $(2+1)$-dimensional dKP hierarchy. In the new variables 
$$
F_{xx}=u, ~ F_{xy}=v, ~ F_{yy}=w,  ~ F_{yt}=s, ~ F_{xt}=p, ~ F_{xz}=uv+s,
$$
the equation (\ref{dkp02}) assumes the quasilinear form
\begin{equation}
\begin{array}{c}
u_y=v_x, ~~ v_y=w_x, ~~ v_t=s_x, ~~ w_t=s_y, ~~ u_t=p_x, ~~ u_z=(uv+s)_x, \\
\ \\
 v_t=p_y, ~~ v_z=(uv+s)_y, ~~ s_x=p_y, ~~ p_z=(uv+s)_t.
\end{array}
\label{dkp12}
\end{equation}
Looking for reductions in the form $u=u(R^1, ..., R^n), $ $v=v(R^1, ..., R^n), $ $ w=w(R^1, ..., R^n), $ $ s=s(R^1, ..., R^n), $ $ p=p(R^1, ..., R^n), $
where the Riemann invariants $R^i$ satisfy  (\ref{R}), and substituting into (\ref{dkp12}),  one readily obtains
\begin{equation}
\partial_i v=\mu^i \partial_i u, ~~~ \partial_i s=\lambda^i \mu^i \partial_i u, ~~~ \partial_i w=(\mu^i)^2 \partial_i u, ~~~ \partial_i p=\lambda^i \partial_i u,
\label{dkp22}
\end{equation}
along with the dispersion relation
\begin{equation}
\eta^i=v+u\mu^i+\mu^i \lambda^i.
\label{dispdkp2}
\end{equation}
The compatibility condition
$\partial_i\partial_jv=\partial_j\partial_iv$ implies
\begin{equation}
\partial_i\partial_ju=\frac{\partial_j\mu^i}{\mu^j-\mu^i}\partial_iu+\frac{\partial_i\mu^j}{\mu^i-\mu^j}\partial_ju,
\label{dkp32}
\end{equation}
while the commutativity condition (\ref{comm}) results in
\begin{equation}
\partial_j\mu^i=\frac{\mu^j+\mu^i}{\lambda^j-\lambda^i}\ \partial_j u, ~~~ \partial_j\lambda^i=\frac{\mu^j+\mu^i}{\mu^j-\mu^i}\ \partial_j u.
\label{dkp42}
\end{equation}
The substitution of (\ref{dkp42}) into (\ref{dkp32}) implies the
over-determined system for  $\mu^i(R),  \lambda^i(R)$ and $u(R)$,
\begin{equation}
\begin{array}{c}
\partial_j\mu^i=\frac{\mu^j+\mu^i}{\lambda^j-\lambda^i}\ \partial_j u, ~~~ \partial_j\lambda^i=\frac{\mu^j+\mu^i}{\mu^j-\mu^i}\ \partial_j u, \\
\ \\
\partial_i\partial_ju=2\frac{\mu^j+\mu^i}{(\mu^j-\mu^i)(\lambda^j-\lambda^i)}\ \partial_iu\partial_ju,
\end{array}
\label{dkp52}
\end{equation}
$i\ne j$, which is analogous to the system (\ref{h6}). There is  one crucial difference: the system ({\ref{dkp52}) is {\it not} in involution. Calculating compatibility conditions
$\partial_k(\partial_j\mu^i)-\partial_j(\partial_k\mu^i)=0$, $\partial_k(\partial_j\lambda^i)-\partial_j(\partial_k\lambda^i)=0$  and
$\partial_k(\partial_j\partial_i u)-\partial_j(\partial_k\partial_iu)=0$, one arrives at  extra relations
\begin{equation}
\begin{array}{c}
\left((\mu^i)^2(\lambda^k-\lambda^j)+(\mu^k)^2(\lambda^j-\lambda^i)+(\mu^j)^2(\lambda^i-\lambda^k)  \right) \ \times \\
\ \\
\left(\mu^k\mu^j(\lambda^k-\lambda^j)+\mu^i\mu^j(\lambda^j-\lambda^i)+\mu^i\mu^k(\lambda^i-\lambda^k)  \right)=0.
\end{array}
\label{bracket}
\end{equation}
Thus, there are two cases to consider.

\noindent {\bf Case 1.} Equating to zero the first bracket in (\ref{bracket}),
$$
(\mu^i)^2(\lambda^k-\lambda^j)+(\mu^k)^2(\lambda^j-\lambda^i)+(\mu^j)^2(\lambda^i-\lambda^k)=0,
$$
one obtains $\lambda^i=a(\mu^i)^2+b$. The substitution of this ansatz   into  (\ref{dkp52})  implies $(\mu^i)^2\partial_ja+\partial_jb=\partial_ju$, so that
$a=\alpha$, $b=u+\beta$ where $\alpha, \beta$ are arbitrary constants. Thus, $\lambda^i=\alpha(\mu^i)^2+u+\beta$. Substituting  $\lambda^i$ into the equation
$ \partial_i p=\lambda^i \partial_i u$ and using  (\ref{dkp22}) one obtains, upon elementary integration, $p=\frac{1}{2}u^2+\beta u+\alpha w+\gamma$. Expressed in terms of second derivatives of $F$, this
constraint reads
$$
F_{xt}=\frac{1}{2}F_{xx}^2+\beta F_{xx}+\alpha F_{yy}+\gamma.
$$
One can show that the constants $\alpha, \beta, \gamma$ are not essentials and can be reduced to $\alpha=1, \beta=\gamma=0$. Thus, the method of hydrodynamic reductions applied to the PDE (\ref{dkp02}) reconstructs the first flow of the dKP hierarchy, $F_{xt}=\frac{1}{2}F_{xx}^2+F_{yy}$.

\noindent {\bf Case 2.} Equating to zero the second bracket in (\ref{bracket}),
$$
\mu^k\mu^j(\lambda^k-\lambda^j)+\mu^i\mu^j(\lambda^j-\lambda^i)+\mu^i\mu^k(\lambda^i-\lambda^k)=0, 
$$
one obtains $\lambda^i=\frac{a}{\mu^i}+b$. The substitution of this ansatz   into  (\ref{dkp52}) implies $\partial_ja+\mu^j\partial_ju+\mu^i(\partial_jb+\partial_ju)=0$, so that
$a=-v+\alpha$, $b=-u+\beta$ where $\alpha, \beta$ are arbitrary constants. Thus, $\lambda^i=\frac{\alpha-v}{\mu^i}+\beta-u$. Substituting  $\lambda^i$ into the equation
$ \partial_i s=\lambda^i \mu^i \partial_i u$ and using  (\ref{dkp22}) one has, upon elementary integration, $s=-uv+\alpha u+\beta v+\gamma$. Expressed in terms of second derivatives of $F$, this
constraint reads
$$
F_{xx}F_{xy}+F_{yt}=\alpha F_{xx}+\beta F_{xy}+\gamma.
$$
Again, the constants $\alpha, \beta, \gamma$ are not essentials and can be reduced to zero. The resulting constraint $F_{yt}+F_{xx}F_{xy}=0$ characterizes stationary points of the flow
(\ref{dkp02}). We have checked that  this constraint is integrable in the sense of hydrodynamic reductions (as a $(2+1)$-dimensional PDE).

In any case, we conclude that the  equation (\ref{dkp02}) is {\it not} integrable as a four-dimensional PDE. This is manifested by the fact that the system (\ref{dkp52}), which governs hydrodynamic reductions, is not in involution.

{\bf Example 4}.  Let us consider the system
\begin{equation}
m_t=n_x, ~~~~ n_z=m_y+mn_x-nm_x,
\label{max}
\end{equation}
which, in the limit $z=t$,  has been extensively investigated in \cite{Pavlov, dun3, Alonso}. Looking for reductions in the form $m=m(R^1, ..., R^n), \ n=n(R^1, ..., R^n)$
where the Riemann invariants $R^i$ satisfy  (\ref{R}), one obtains
$$
\partial_i n=\lambda^i \partial_im, ~~~ \mu^i=\lambda^i \eta^i-m\lambda^i+n.
$$
The commutativity conditions (\ref{comm}) imply
$$
\partial_j\eta^i=\partial_jm, ~~~ \partial_i\partial_jm=0,
$$
hence, up to  reparametrizations $R^i\to \varphi^i(R^i)$, one has
$$
m=\sum_k R^k, ~~~ \eta^i=f^i(R^i)+\sum_k R^k, 
$$
where $f^i(R^i)$  are arbitrary functions of a
single variable.  The characteristic speeds $\lambda^i$ solve the linear system
$$
\frac{\partial_j \lambda^i}{\lambda^j-\lambda^i}=\frac{1}{f^j(R^j)-f^i(R^i)}.
$$
We refer to \cite{Pavlov} for the general formula for $\lambda^i$ and further discussion of this example 
in the $(2+1)$-dimensional limit  $z=t$.
Thus, $n$-component hydrodynamic reductions of the system (\ref{max}) are parametrized by 
$2n$ arbitrary functions of one variable ($n$ functions $f^i(R^i)$ plus $n$ functions in the general solution of the linear system for $\lambda^i$). Therefore, the four-dimensional system (\ref{max}) is integrable. Notice that it can be obtained as the condition of commutativity  of two vector fields,
$$
[\partial_z-m\partial_x-\lambda \partial_x, ~~ \partial_y-n\partial_x-\lambda \partial_t]=0,
$$
$\lambda=const$, compare with \cite{dun3}. Some further multi-dimensional generalizations of this example are discussed in Sect.3.

{\bf Example 5.} The six-dimensional generalization of the heavenly equation,
\begin{equation}
\theta_{t\tilde t}+\theta_{z\tilde z}+\theta_{tx}\theta_{zy}-\theta_{ty}\theta_{zx}=0,
\label{P1}
\end{equation}
has been proposed in   \cite{Przanovski}. Introducing  the variables $\theta_{tx}=a, $ $ \theta_{zy}=b, $ 
$ \theta_{ty}=p, $ $ \theta_{zx}=q, $ $\theta_{z\tilde z}=r,$ $ \theta_{t\tilde t}=pq-ab-r$, one can rewrite 
(\ref{P1}) in a quasilinear form,
\begin{equation}
\begin{array}{c}
a_y=p_x, ~~~ a_z=q_t, ~~~ b_t=p_z, ~~~ b_x=q_y, ~~~ b_{\tilde z}=r_y, ~~~ q_{\tilde z}=r_x,\\
\ \\
 p_{\tilde z}=(pq-ab-r)_y.
\end{array}
\label{P2}
\end{equation}
Hydrodynamic reductions are sought in the form $a=a(R^1, ..., R^n),$ $b=b(R^1, ..., R^n),$ $p=p(R^1, ..., R^n),$ $q=q(R^1, ..., R^n), $  $r=r(R^1, ..., R^n), $ where the Riemann invariants $R^1, ..., R^n$  solve the commuting equations
$$
R^i_{x}=\lambda^i(R)\ R^i_{z}, ~~~~ R^i_{ y}=\mu^i(R)\ R^i_{z}, ~~~~ R^i_{\tilde z}=\eta^i(R)\ R^i_{z}, ~~~
R^i_{t}=\beta^i(R)\ R^i_{z}, ~~~ R^i_{\tilde t}=\gamma^i(R)\ R^i_{z}.
$$
The substitution into (\ref{P2}) implies
\begin{equation}
\partial_i p=\beta^i\partial_i b, ~~~ \partial_i r=\frac{\eta^i}{\mu^i}\partial_i b, ~~~ \partial_iq=\frac{\lambda^i}{\mu^i}\partial_i b, ~~~ \partial_i a=\frac{\lambda^i \beta^i}{\mu^i}\partial_i b,
\label{P4}
\end{equation}
along with the dispersion relation
\begin{equation}
\eta^i=\beta^i\mu^i q+\lambda^i p-\beta^i\lambda^i b-\mu^i a-\beta^i\gamma^i.
\label{P5}
\end{equation}
Substituting  $\lambda^i$ into the commutativity conditions
$$
\frac{\partial_j\lambda
^i}{\lambda^j-\lambda^i}=\frac{\partial_j\mu^i}{\mu^j-\mu^i}=\frac{\partial_j\eta^i}{\eta^j-\eta^i}=
\frac{\partial_j\beta^i}{\beta^j-\beta^i}=\frac{\partial_j\gamma^i}{\gamma^j-\gamma^i},
$$
and taking into account that the compatibility conditions for the relations $\partial_i p=\beta^i\partial_i b$ imply
$$
\partial_i\partial_jb=\frac{\partial_j\beta^i}{\beta^j-\beta^i}\partial_ib+\frac{\partial_i\beta^j}{\beta^i-\beta^j}\partial_jb,
$$
one arrives at the following system:
\begin{equation}
\begin{array}{c}
\frac{\partial_j\beta^i}{\beta^j-\beta^i}=\frac{\partial_j\lambda^i}{\lambda^j-\lambda^i}=
\frac{\partial_j\mu^i}{\mu^j-\mu^i}=\frac{\partial_j\gamma^i}{\gamma^j-\gamma^i}=
\frac{\lambda^i-\lambda^j\mu^i/\mu^j}{q(\mu^j-\mu^i)+b(\lambda^i-\lambda^j)+\gamma^i-\gamma^j}
\ \partial_jb, \\
\ \\
\partial_i\partial_jb=
\frac{\lambda^i(1+\mu^j/\mu^i)-\lambda^j(1+\mu^i/\mu^j)}{q(\mu^j-\mu^i)+b(\lambda^i-\lambda^j)+\gamma^i-\gamma^j}\ \partial_ib\partial_jb.
\end{array}
\label{P6}
\end{equation}
Solving   equations (\ref{P6}) for $\beta^i$, $\lambda^i$, $\mu^i$,  $\gamma^i$ and $b$, determining $\eta^i$ from the dispersion relation (\ref{P5}) and calculating $p, r, q, a$ from the equations (\ref{P4}) 
(which are automatically compatible by virtue of (\ref{P6})), one obtains the general $n$-component  reduction of the  equation (\ref{P1}).
The commutativity conditions  will  be satisfied identically.
We have checked that the system (\ref{P6}) is in involution and its general solution depends, up to
reparametrizations $R^i\to \varphi^i(R^i)$,  on $4n$ arbitrary functions of one variable.

\section{Multi-dimensional linearly degenerate  systems of hydrodynamic type}

In the recent publication \cite{Fer4}, we gave a complete characterization of two-component $(2+1)$-dimensional integrable systems of hydrodynamic type,
$$
\left(\begin{array}{c}
v\\
w
\end{array}\right)_t+A(v, w)\left(\begin{array}{c}
v\\
w
\end{array}\right)_x+B(v, w)\left(\begin{array}{c}
v\\
w
\end{array}\right)_y=0,
$$
which possess infinitely many hydrodynamic reductions. The integrability conditions constitute a complicated overdetermined system of second order PDEs for $2\times 2$ matrices $A$ and $B$.
In the particular case when the matrix $A$ is assumed to be 
linearly degenerate,
\begin{equation}
 A=\left(
\begin{array}{cc}
w & 0 \\
\ \\
0 & v
\end{array}
\right),
 \label{lindeg}
 \end{equation}
 these  conditions imply
 $$
 B=\left(
\begin{array}{cc} 
 \frac{f(w)}{w-v}-\alpha w^2 &  \frac{f(v)}{w-v} \\ 
 \ \\
\frac{f(w)}{v-w} & \frac{f(v)}{v-w}-\alpha v^2
 \end{array}
 \right)
$$
where $f$ is a cubic polynomial, $f(v)=\alpha v^3+\beta v^2 +\gamma v 
+\delta$, and $\alpha, \beta, \gamma, \delta$ are arbitrary 
constants. A remarkable property of this example is that
{\it any} matrix in the linear pencil $B+\mu A$ is also linearly 
degenerate (that is, reduces to the diagonal form (\ref{lindeg}) after  an appropriate change of dependent variables). Explicitly, one has
 $$
 \begin{array}{c}
  B=\delta B_1+\gamma B_2+\beta B_3+\alpha B_4= \\
 \ \\
 \delta \left(
\begin{array}{cc} 
 \frac{1}{w-v} &  \frac{1}{w-v} \\ 
\frac{1}{v-w} & \frac{1}{v-w}
 \end{array} \right)+
 \gamma \left(
\begin{array}{cc} 
 \frac{w}{w-v} &  \frac{v}{w-v} \\ 
\frac{w}{v-w} & \frac{v}{v-w}
 \end{array}
 \right)+
 \beta \left(
\begin{array}{cc} 
 \frac{w^2}{w-v} &  \frac{v^2}{w-v} \\ 
\frac{w^2}{v-w} & \frac{v^2}{v-w}
 \end{array}
 \right)+
 \alpha \left(
\begin{array}{cc} 
 \frac{vw^2}{w-v} &  \frac{v^3}{w-v} \\ 
\frac{w^3}{v-w} & \frac{wv^2}{v-w}
 \end{array}
 \right).
 \end{array}
$$
Let us introduce the $(5+1)$-dimensional system
$$
\left(\begin{array}{c}
v\\
w
\end{array}\right)_t+A\left(\begin{array}{c}
v\\
w
\end{array}\right)_x+ \\
\ \\
B_1\left(\begin{array}{c}
v\\
w
\end{array}\right)_y+
B_2\left(\begin{array}{c}
v\\
w
\end{array}\right)_z+
B_3\left(\begin{array}{c}
v\\
w
\end{array}\right)_s+
B_4\left(\begin{array}{c}
v\\
w
\end{array}\right)_r
=0.
$$
Notice that an arbitrary linear combination of matrices $A$ and $B_1, B_2, B_3, B_4$ is linearly degenerate!  In the new variables $m=v+w,\  n=vw$, this systems reduces to
$$
m_t+n_x+nm_r-mn_r=0, ~~~
n_t+mn_x-nm_x+m_y+n_z+mn_s-nm_s=0,
$$
taking a fully symmetric form
\begin{equation}
m_{\tilde t}=n_{\tilde x}+nm_{\tilde r}-mn_{\tilde r},  ~~~~ n_{\tilde z}=m_{\tilde y}+mn_{\tilde s}-nm_{\tilde s},
\label{max1}
\end{equation}
 after the obvious  linear change of independent variables. In the limit ${\tilde s}=\tilde x, \ m_{\tilde r}=n_{\tilde r}=0$ it reduces to the system (\ref{max}) from Example 4. 
Notice that the system (\ref{max1}) arises as  the condition of commutativity  of two vector fields,
$$
[\partial_ {\tilde z}-m\partial_{\tilde s}-\lambda \partial_{\tilde x}+\lambda m \partial_{\tilde r}, ~~
\partial_{\tilde y}-n\partial_{\tilde s}-\lambda \partial_{\tilde t}+\lambda n \partial_{\tilde r}]=0.
$$
Let us demonstrate that the system (\ref{max1}) possesses `enough' hydrodynamic reductions and,
therefore, should be regarded as an integrable systems in $5+1$ dimensions. Looking for reductions in the form $m=m(R^1, ..., R^n), \ n=n(R^1, ..., R^n)$ where the Riemann invariants $R^i$ solve five commuting systems
$$
R^i_{\tilde t}=\lambda^i(R)\ R^i_{\tilde x}, ~~~~ R^i_{\tilde y}=\mu^i(R)\ R^i_{\tilde x}, ~~~~ R^i_{\tilde z}=\eta^i(R)\ R^i_{\tilde x}, ~~~
R^i_{\tilde r}=\beta^i(R)\ R^i_{\tilde x}, ~~~ R^i_{\tilde s}=\gamma^i(R)\ R^i_{\tilde x},
$$
and  substituting into (\ref{max1}), we obtain
$$
(\lambda^i-n\beta^i) \partial_i m=(1-m\beta^i)\partial_in, ~~~
(\eta^i-m\gamma^i) \partial_i n=(\mu^i-n\gamma^i)\partial_im.
$$
Setting $\partial_in=\varphi^i\partial_im$, one obtains  expressions for $\lambda^i$ and $\mu^i$ 
in the form
$$
\lambda^i=n\beta^i+(1-m\beta^i)\varphi^i, ~~~
\mu^i=n\gamma^i+(\eta^i-m\gamma^i)\varphi^i,
$$
as well as the consistency condition
$$
\partial_i\partial_jm=\frac{\partial_j\varphi^i}{\varphi^j-\varphi^i}\partial_im+\frac{\partial_i\varphi^j}{\varphi^i-\varphi^j}\partial_jm.
$$
Inserting the expressions for $\lambda^i$ and $\mu^i$ into the commutativity conditions
$$
\frac{\partial_j\lambda
^i}{\lambda^j-\lambda^i}=\frac{\partial_j\mu^i}{\mu^j-\mu^i}=\frac{\partial_j\eta^i}{\eta^j-\eta^i}=
\frac{\partial_j\beta^i}{\beta^j-\beta^i}=\frac{\partial_j\gamma^i}{\gamma^j-\gamma^i},
$$
one ends up with the following equations for $\eta^i, \beta^i, \gamma^i, \varphi^i$ and $m$:
\begin{equation}
\begin{array}{c}
\frac{\partial_j\eta^i}{\eta^j-\eta^i}=
\frac{\partial_j\beta^i}{\beta^j-\beta^i}=\frac{\partial_j\gamma^i}{\gamma^j-\gamma^i}=
\frac{\beta^i\eta^i-\gamma^i}{(1-m\beta^j)(\eta^i-m\gamma^i)-(1-m\beta^i)(\eta^j-m\gamma^j)}
\ \partial_jm, \\
\ \\
\frac{\partial_j\varphi^i}{\varphi^j-\varphi^i}=
\frac{\beta^i\eta^j-\gamma^i+m(\gamma^i\beta^j-\gamma^j\beta^i)}{(1-m\beta^j)(\eta^i-m\gamma^i)-(1-m\beta^i)(\eta^j-m\gamma^j)}\ \partial_jm, \\
\ \\
\partial_i\partial_jm=
\frac{\gamma^j-\gamma^i+\beta^i\eta^j-\beta^j\eta^i+2m(\gamma^i\beta^j-\gamma^j\beta^i)}{(1-m\beta^j)(\eta^i-m\gamma^i)-(1-m\beta^i)(\eta^j-m\gamma^j)}\ \partial_im\partial_jm.
\end{array}
\label{eq}
\end{equation}
It has been verified directly that this system is in involution and its general solution depends, modulo reparametrizations $R^i\to f^i(R^i)$, on  $4n$ arbitrary functions of one variable, thus manifesting the integrability of the $(5+1)$-dimensional system (\ref{max1}).

\section{Appendix}
Here we  apply the method of hydrodynamic reductions to the classification of integrable PDEs of the form
\begin{equation}
\theta_{tx}+\theta_{zy}=f(\theta_{xx}, \theta_{xy}, \theta_{yy}).
\label{A1}
\end{equation}
Introducing new variables $\theta_{xx}=u, $ $ \theta_{xy}=v, $ $ \theta_{yy}=w, $ $ \theta_{tx}=p, $
 $ \theta_{zy}=f(u, v, w)-p$, one rewrites 
(\ref{A1}) in the quasilinear form
\begin{equation}
\begin{array}{c}
u_y=v_x, ~~~ u_t=p_x, ~~~ v_y=w_x, ~~~ v_t=p_y, \\
\ \\
v_z=(f(u, v, w)-p)_x, ~~~ w_z=(f(u, v, w)-p)_y.
\end{array}
\label{A2}
\end{equation}
Hydrodynamic reductions are sought in the form $u=u(R^1, ..., R^n),$ $v=v(R^1, ..., R^n),$ $w=w(R^1, ..., R^n),$ $p=p(R^1, ..., R^n)$  where the Riemann invariants $R^1, ..., R^n$  solve a triple of commuting hydrodynamic type systems (\ref{R}).
The substitution into (\ref{A2}) implies 
\begin{equation}
\partial_i p=\lambda^i\partial_i u, ~~~ \partial_i v=\mu^i\partial_i u, ~~~ \partial_i w=(\mu^i)^2\partial_i u,
\label{A4}
\end{equation}
along with the dispersion relation
\begin{equation}
\lambda^i=f_u+f_v\mu^i+f_w(\mu^i)^2-\mu^i\eta^i.
\label{A5}
\end{equation}
Substituting  $\lambda^i$ into the commutativity conditions (\ref{comm}),
and taking into account that the compatibility conditions for the relations (\ref{A4})  imply
$$
\partial_i\partial_ju=\frac{\partial_j\mu^i}{\mu^j-\mu^i}\partial_iu+\frac{\partial_i\mu^j}{\mu^i-\mu^j}\partial_ju,
$$
one arrives at the following system:
\begin{equation}
\begin{array}{c}
\partial_j\mu^i=\frac{S_{ij}}{f_w(\mu^j-\mu^i)+\eta^i-\eta^j} \ \partial_ju, ~~~~
\partial_j\eta^i=\frac{\eta^j-\eta^i}{\mu^j-\mu^i}\frac{S_{ij}}{f_w(\mu^j-\mu^i)+\eta^i-\eta^j} \ \partial_ju, \\
\ \\
\partial_i\partial_j u=\frac{2}{\mu^j-\mu^i}\frac{S_{ij}}{f_w(\mu^j-\mu^i)+\eta^i-\eta^j} \ \partial_iu\partial_ju;
\end{array}
\label{A6}
\end{equation}
here
$$
\begin{array}{c}
S_{ij}=f_{uu}+(\mu^i+\mu^j)f_{uv}+((\mu^i)^2+(\mu^j)^2)f_{uw}+\mu^i\mu^j f_{vv}+ \\
\mu^i\mu^j(\mu^i+\mu^j)f_{vw}+(\mu^i)^2(\mu^j)^2f_{ww}.
\end{array}
$$
Compatibility conditions  for the system (\ref{A6}) are of the form
$$
\begin{array}{c}
\partial_k (\partial_j\mu^i)-\partial_j (\partial_k\mu^i)=(...)\ \partial_j u \partial_k u, ~~~
\partial_k (\partial_j\eta^i)-\partial_j (\partial_k\eta^i)=(...)\ \partial_j u \partial_k u,\\
\ \\
\partial_k (\partial_j\partial_i u)-\partial_j (\partial_k\partial_i u)=(...)\ \partial_j u \partial_k u,
\end{array}
$$
where dots $(...)$ denote complicated {\it rational} expressions in $\mu^i, \mu^j, \mu^k$ and 
$\eta^i, \eta^j, \eta^k$ whose coefficients are functions of the derivatives of $f$ up to third order. Equating these rational expressions to zero one arrives at the following system  for $f$:
$$
f_{uu}=f_{ww}=f_{uv}=f_{wv}=f_{vv}-2f_{uw}=f_{vvv}=0.
$$
Up to elementary changes of variables,
the general nonlinear solution of this system corresponds to the second heavenly equation (\ref{h1}).

\section{Concluding remarks}

We have demonstrated that the requirement of  existence of `sufficiently many' $n$-component reductions can be used as the effective criterion providing the test for integrability  of multi-dimensional dispersionless PDEs. We believe that using the approach outlined in this paper along with the available computer algebra packages (we have used Mathematica 5.0), one can obtain complete lists of multi-dimensional integrable systems within various particularly interesting classes, hyperbolic systems of hydrodynamic type being one of them. Partial classification results can be found in \cite{Fer3, Fer4, Pavlov1}. The main problems arising here are the complexity of integrability conditions (making difficult their geometric analysis), and the volume of symbolic calculations required. 

\section*{Acknowledgements}

We would like to thank  M. Dunajski and M. Pavlov  for stimulating discussions and references.

\end{document}